\documentclass[9.5pt,journal,finalsubmission,compsoc]{IEEEtran}

\ifCLASSOPTIONcompsoc
  \usepackage[nocompress]{cite}
\else
  \usepackage{cite}
\fi

\usepackage{amsmath,amsfonts}
\usepackage{blindtext}
\usepackage{algorithm}
\usepackage{algpseudocode}
\usepackage{array}
\usepackage[caption=false,font=normalsize,labelfont=sf,textfont=sf]{subfig}
\usepackage{textcomp}
\usepackage{stfloats}
\usepackage{url}
\usepackage{verbatim}
\usepackage{graphicx}
\usepackage{multirow}
\usepackage{booktabs}
\usepackage{amsbsy}
\newcolumntype{C}[1]{>{\centering\let\newline\\\arraybackslash\hspace{0pt}}m{#1}}

\title{Polytope: An Algorithm for Efficient Feature Extraction on Hypercubes}
\author{Mathilde Leuridan, James Hawkes, Simon Smart, Emanuele Danovaro and Tiago Quintino
\thanks{
All authors are with the European Center for Medium-Range Weather Forecasts (ECMWF).
Contact e-mail: mathilde.leuridan@ecmwf.int
}
}

\IEEEtitleabstractindextext{
\begin{abstract}
Data extraction algorithms on data hypercubes, or datacubes, are traditionally only capable of cutting boxes of data along the datacube axes. For many use cases however, this is not a sufficient approach and returns more data than users might actually need. This not only forces users to apply post-processing after extraction, but more importantly this consumes more I/O resources than is necessary. When considering very large datacubes from which users only want to extract small non-rectangular subsets, the box approach does not scale well. Indeed, with this traditional approach, I/O systems quickly reach capacity, trying to read and return unwanted data to users. In this paper, we propose a novel technique, based on computational geometry concepts, which instead carefully pre-selects the precise bytes of data which the user needs in order to then only read those from the datacube. As we discuss later on, this novel extraction method will considerably help scale access to large petabyte size data hypercubes in a variety of scientific fields. \\
\end{abstract}
}

\begin{document}

\maketitle
\IEEEdisplaynontitleabstractindextext
\IEEEpeerreviewmaketitle

\ifCLASSOPTIONcompsoc
\IEEEraisesectionheading{\section{Introduction}\label{sec:introduction}}
\else
\section{Introduction}

\label{sec:introduction}
\fi
\IEEEPARstart{I}{n}
the past century, fields  in science and technology have entered a new era - the era of ``big data''. From weather forecasting to medicine, scientific advances have led to a surge in the quantity of data produced daily. 
Indeed, scientific data has been steadily growing in the past decades and in recent years especially, it has experienced exponential growth.
Whilst this new era holds many promises for major scientific developments in the years to come, the question arises of how to efficiently use this wealth of data.\\
The scientific data collected nowadays often depends on a number of different variables and can thus be represented as a multidimensional array, or datacube \cite{datacube}. Organising data inside such datacubes has attracted a lot of interest in the past few years, with many tools now available to work on such data representations. Most modern software architectures provide support to handle such data structures, from Matlab \cite{matlab} to Python \cite{xarray} and C++ \cite{xtensor}.
However, in each of these software, datacubes can only access data ``orthogonally'' to their ``axes" by selecting specific values or ranges along given dimensions \cite{datacube,sql,tquel,olap}. \\
Such limited data access mechanisms in the form of bounding boxes are non-optimal for a wide range of applications. 
Consider for example the case where a user wants to access temperature data over a country. For this particular example, the bounding box data extraction approach proves to be quite inconvenient as country shapes are not well-represented by bounding boxes.
This then not only implies that much more data than is necessary is read and returned from the datacube, thereby consuming more I/O resources, but it also places the additional burden of post-processing on the user after retrieval.\\
\noindent
To address this issue, we introduce a new alternative way of accessing datacubes. Our extraction algorithm, Polytope\footnote{\url{https://github.com/ecmwf/polytope}}, enables users to efficiently query arbitrary high-dimensional shapes from a datacube, slicing non-orthogonally along the datacube's axes. This is much less restrictive than the popular bounding box approach described above and constitutes a major improvement compared to existing data extraction methods. \\
Indeed, as discussed above in the country slicing example, traditional bounding-box extraction methods are insufficient for handling such complicated requests. The Polytope algorithm however was designed especially with these requests in mind and is able to directly extract such shapes from very large data hypercubes. Because our algorithm computes the exact bytes that users are interested in and only reads those from the datacube, it scales well to large high-dimensional request shapes unlike the traditional bounding-box extraction techniques which scale with the tensor product of each dimension. The Polytope algorithm will thus enable scientists to efficiently make use of their ever increasing data, whilst improving the efficiency of their I/O system.\\
In this paper, we first introduce the idea behind the Polytope algorithm, before describing its inner mechanism in detail. We then expose some of its possible applications in different scientific fields before finally performing a first analysis of the algorithm's performance.
\section{Concept}
\begin{figure}[!ht]
         \centering
         \includegraphics[width=    1\linewidth]{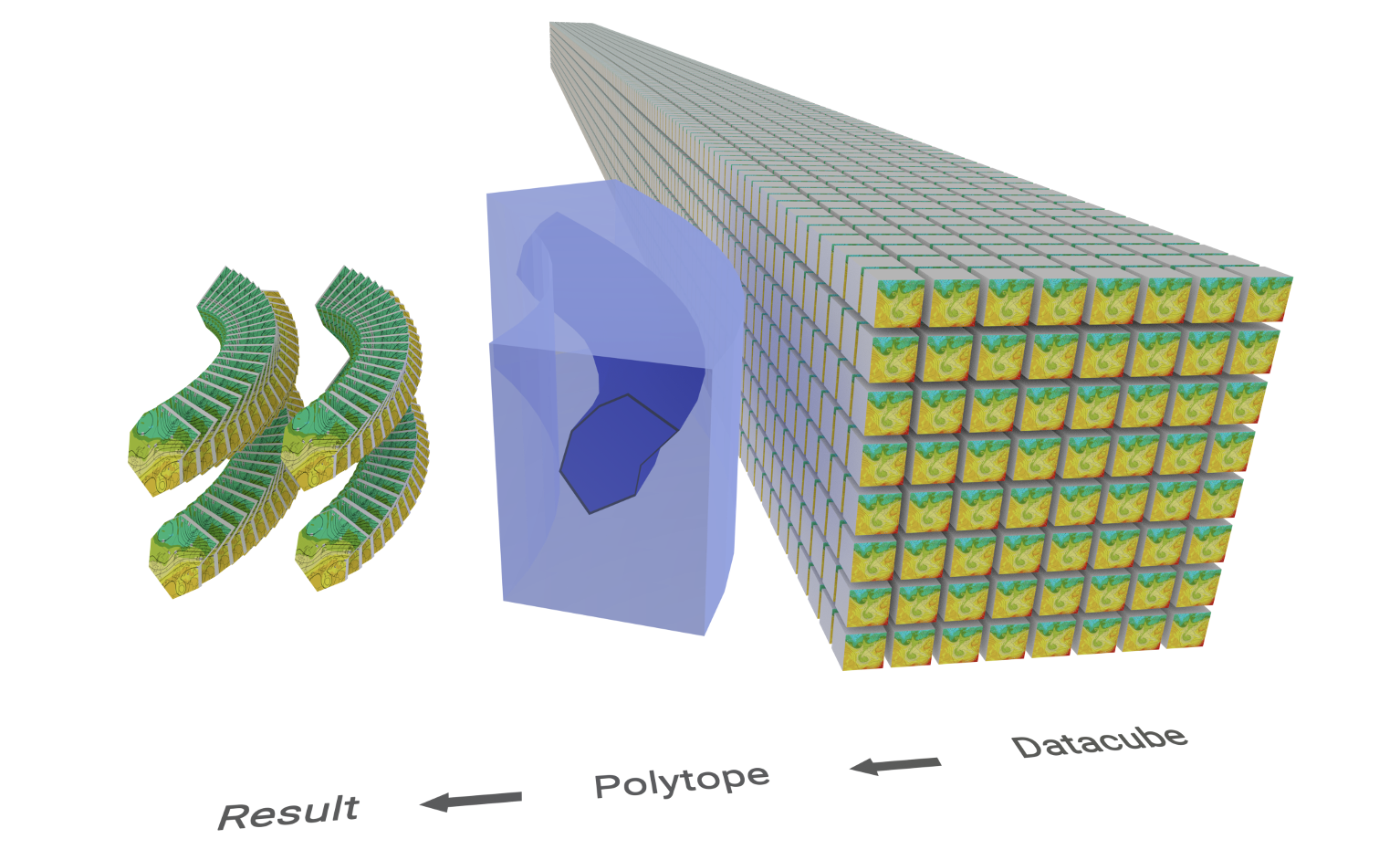}
         \caption{Concept of the Polytope algorithm. A 6-dimensional polytope stencil is used to extract data of this shape from the datacube.}
         \label{Polytope_concept}
\end{figure}

Before diving into a technical description of our software, let us first explain in more detail the conceptual approach we take. 
With the Polytope algorithm, we developed a data extraction algorithm which supports the retrieval of arbitrary high-dimensional request shapes, called \emph{features}, from arbitrary data hypercubes. Our algorithm is not restricted to any particular request shapes or application field and is in fact intended to be generic and work seamlessly in any scientific application involving datacubes.\\
Rather than pre-defining a set of shapes which can be extracted from the datacube, the Polytope algorithm takes n-dimensional \emph{polytope} shapes as input, giving the algorithm its name. In computational geometry, a polytope is defined as the convex hull of a given point set $\mathcal{P} = \{p_1, \dots, p_n\}$ \cite{polytopedef}.
\noindent
Note that polytopes are convex by definition. Polytopes can in fact be thought of as high-dimensional convex polygons. 
In the Polytope software, we use polytopes because any arbitrary high-dimensional shape, even a concave shape, can be either approximated by or decomposed into simpler convex polytopes. Indeed, polytopes can be seen as the building blocks of high-dimensional geometry. They form the basis of most modern meshing softwares and are used daily in computer graphics to model intricate objects \cite{mesh}. We thus see that by formulating data requests as polytopes, users will in theory be able to request almost any feature of interest to them from a datacube.\\
The underlying idea behind the Polytope algorithm can be visualised in Figure \ref{Polytope_concept}.
\section{Polytope Extraction Algorithm}
We now introduce the Polytope feature extraction algorithm, highlighting in particular the way in which it achieves polytope-based feature extraction on datacubes. \\
Note that the Polytope algorithm only works on a subset of datacubes which possess particular properties. We thus first discuss some of these datacube properties before describing the complete mechanism behind the feature extraction algorithm.
\subsection{Datacube}
Datacubes can be thought of as multi-dimensional arrays. 
In particular, they store data points along different datacube dimensions. Each datacube dimension has an associated ``axis'' metadata with a discrete set of indices stored on it. A data point is then located at each of these indices, forming a datacube.
Each datacube structure is unique however and the Polytope datacube component specifies querying mechanisms on each of these different structures. Moreover, it also describes essential features of the underlying datacube, such as its axes. This helps construct a common framework for treating various types of datacube structures. 
\subsubsection*{\textbf{Axes}}
Axes in a datacube refer to the dimensions along which the data is stored. Values along these axes are called indices. In the Polytope extraction algorithm, we differentiate between two main types of axes, the ordered and unordered categorical axes. These two types of axes cannot be treated in the same way within the slicing step of the algorithm, which leads to their distinction here. 
\paragraph*{Ordered Axes} These axes only accept sets of comparable indices which can be ordered. In particular here, this means that values on ordered axes need to be comparable to each other, such that they must meaningfully support comparison operators ($==$, $<$, $\leq$, $>$, $\geq$). This property then directly implies an ordering between indices on ordered axes. Importantly, note that indices on ordered axes do not have to be integers, but can in fact be any countable type that supports a comparison operation, such as time entities, floating point numbers and of course integers. For such axes, it is possible to query ranges of indices as well as individual axis values. 
\paragraph*{Categorical Axes} The other type of axes which can be handled by our algorithm are categorical axes. These axes only support distinct indices which are not comparable to each other, such as string indices for example. In this case, unlike for ordered axes, it does not make sense to query ranges of indices. Instead, the only possible queries on categorical axes are specific index selections. 
 \\[10pt]
\noindent 
Note that, in practice, indices on a datacube will always have some gap between them, even if it is just a small tolerance. This implies that the set of indices on a datacube axis will always be discrete. All ordered axes are thus countable axes, for which indices can be ordered and numbered using natural numbers.
Note also that the indices on ordered axes do not have to be uniformly spaced. In particular, the datacube axes can be irregular and sparse in their indices.
Lastly, observe that ordered axes can exhibit special behaviours, such as cyclicity along their indices. We thus further subdivide the ordered axis class with as many special subclasses as required to capture all possible axis behaviours. \\
All axes within either of these axis classes can be treated in the same fashion. This allows us to take a common approach towards extracting indices on those axes and thus facilitates the data extraction algorithm. 
\subsubsection*{\textbf{Datacube Structure}}
\begin{figure}[!ht]
         \centering
         \includegraphics[width=  1 \linewidth]{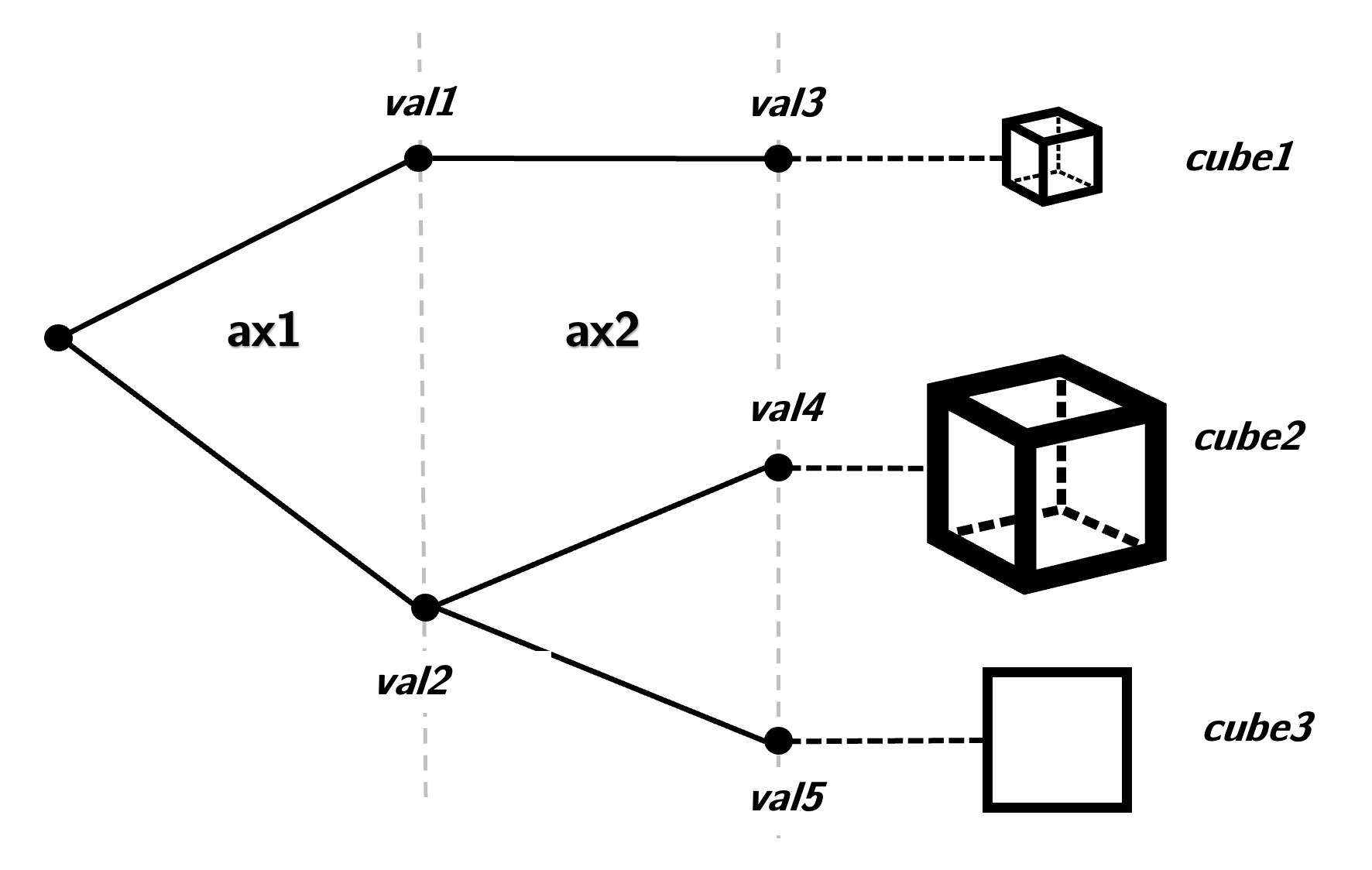}
         \caption{Datacube represented as a non-regular imbalanced tree. Each tree layer represents different axis dimensions with the nodes being corresponding axis index values. Here, the first axis dimension is ax1, which has indices val1 and val2. The second axis dimension ax2 with indices val3, val4 and val5 then branches out onto datacubes of different sizes and dimensions, causing both imbalance and non-regularity in this tree.}
         \label{datacube_tree}
\end{figure}

The datacube can be viewed as a possibly non-regular imbalanced tree. This can be seen in Figure \ref{datacube_tree} and we now explain each of these two datacube properties, non-regularity and imbalance, in more detail with the help of an example.\\
\noindent
Note that the datacube does not necessarily have the same dimensionality in all directions. On some axes, it is possible to have axis indices which give rise to different subsequent axes or axis values. Consider for example the datacube in Figure \ref{datacube_tree} with the \textbf{ax2} axis with indices \textbf{val4} and \textbf{val5}. If we pick index \textbf{val5}, the other axes in the datacube are \pmb{$u$} and \pmb{$v$}, whereas if we pick index \textbf{val4} instead, the other axes in the datacube are \pmb{$x$}, \pmb{$y$} and \pmb{$z$}. This phenomenon can be viewed as a non-regular branching of the datacube axes. This is an important feature of the datacube, which we should take into consideration when thinking about the datacube structure. 
In particular, this suggests that there is a natural ordering of the axes, which we should follow when extracting data. \\
The imbalance in the datacube tree comes from the fact that some datacube axis can possibly have many more indices than others. In our example datacube above, imagine for instance that the \pmb{$u$} and \pmb{$v$} axis each only have 2 index values, whereas the \pmb{$x$}, \pmb{$y$} and \pmb{$z$} axis each have 10 index values. This implies that the \textbf{val4} index has many more children than then \textbf{val5} index and makes this particular datacube very imbalanced. Again, this is a feature of the datacube which is important to remember in order to understand the complete datacube structure.
\subsection{Slicer}
The core of the Polytope feature extraction algorithm is the \emph{slicer}, which contains a novel slicing step on the datacube indices. The slicing algorithm introduced here is of particular relevance, as it supports non-orthogonal slicing across arbitrary ordered axes.
This is in contrast to most state of practice data extraction techniques, which only support range selections on individual axes \cite{sql, tquel}. Indeed, current state of practice data extraction techniques often only cut boxes of data, whereas our slicing algorithm has the capability of cutting polytopes of data. It is also important to note that the slicing algorithm introduced here works on all ordered axes, without any specific constraints about the type of indices stored on these axes. Moreover, as the algorithm is able to handle shapes of arbitrary dimensions, it can be used to extract various low- and high-dimensional queries, making it a highly versatile technique.
\subsubsection*{\textbf{Concept}}
The slicing algorithm used in Polytope differs from others as it is capable of slicing non-orthogonally along datacube axes. By leveraging results in the field of computational geometry, it can extract any convex polytope from the original datacube.
The underlying concept is that we successively slice the requested polytope along each axis in the natural axis ordering using hyperplanes, reducing the dimensionality of the polytope at each step until we are left with a list of all points contained in this polytope. 
\subsubsection*{\textbf{Ordered vs Categorical Axes}}
As mentioned earlier, the slicer handles ordered and categorical axes slightly differently.\\
In particular, categorical axes do not support range queries and thus we can only ask for specific values on these axes, instead of polytopes. For categorical axes, the algorithm therefore only has to check whether the queried indices exist in the datacube, as would happen in every other traditional extraction algorithm. \\
The true innovation of the Polytope extraction technique is its ability to handle arbitrary polytope requests, which it achieves by introducing a new slicing step along the ordered axes. 
Note however that this slicing technique only works on ordered axes for two reasons.\\ 
Firstly, since it is only possible to define and request ranges on ordered axes, it also only makes sense to define polytopes along such axes. Secondly, the slicing step introduced below only works on indices which can be interpolated. As we now explain, these are in fact precisely the ordered axes' indices. Indeed, note that, for the purposes of our algorithm, we assume that all of the ordered axes are measurable and linear axes, which can have continuous index values. We make this assumption even for ordered axes which are only truly countable with gaps between their indices. Because all ordered axes have some comparison operation, this is a valid assumption. This then implies that we can perform interpolation on all of the ordered axes' indices. The slicing step thus works on all ordered axes, but not on the categorical axes.  
\subsubsection*{\textbf{Slicing Step}}
\begin{figure*} 
\captionsetup[subfigure]{font=footnotesize}
    \centering
  \subfloat[2D example\label{2d}]{
       \includegraphics[width=0.45\linewidth]{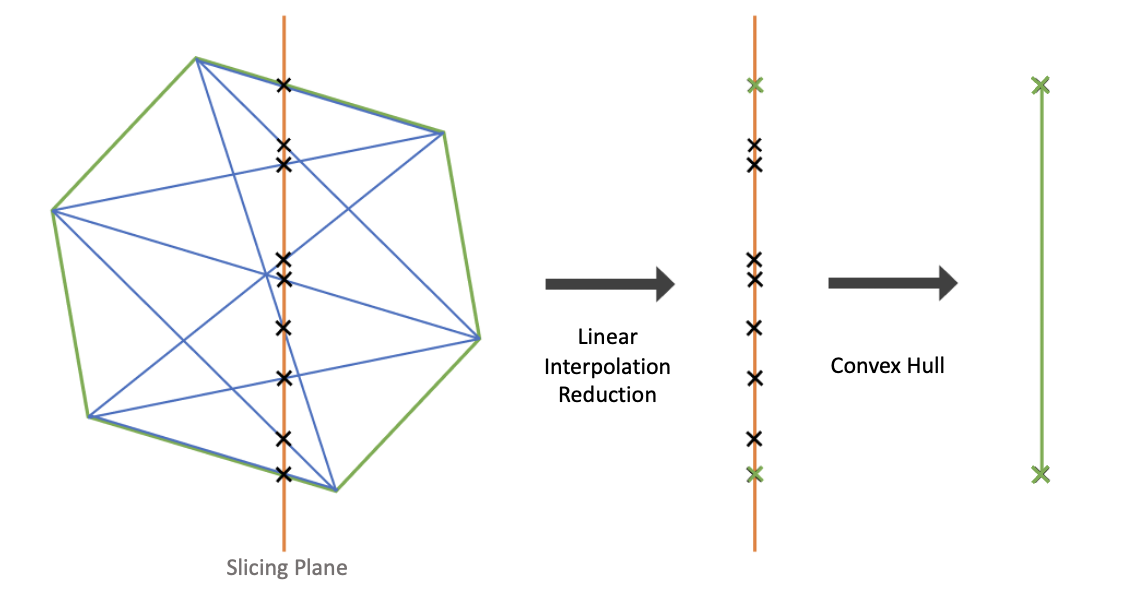}}
    \hfill
  \subfloat[3D example\label{3d}]{
        \includegraphics[width=0.45\linewidth]{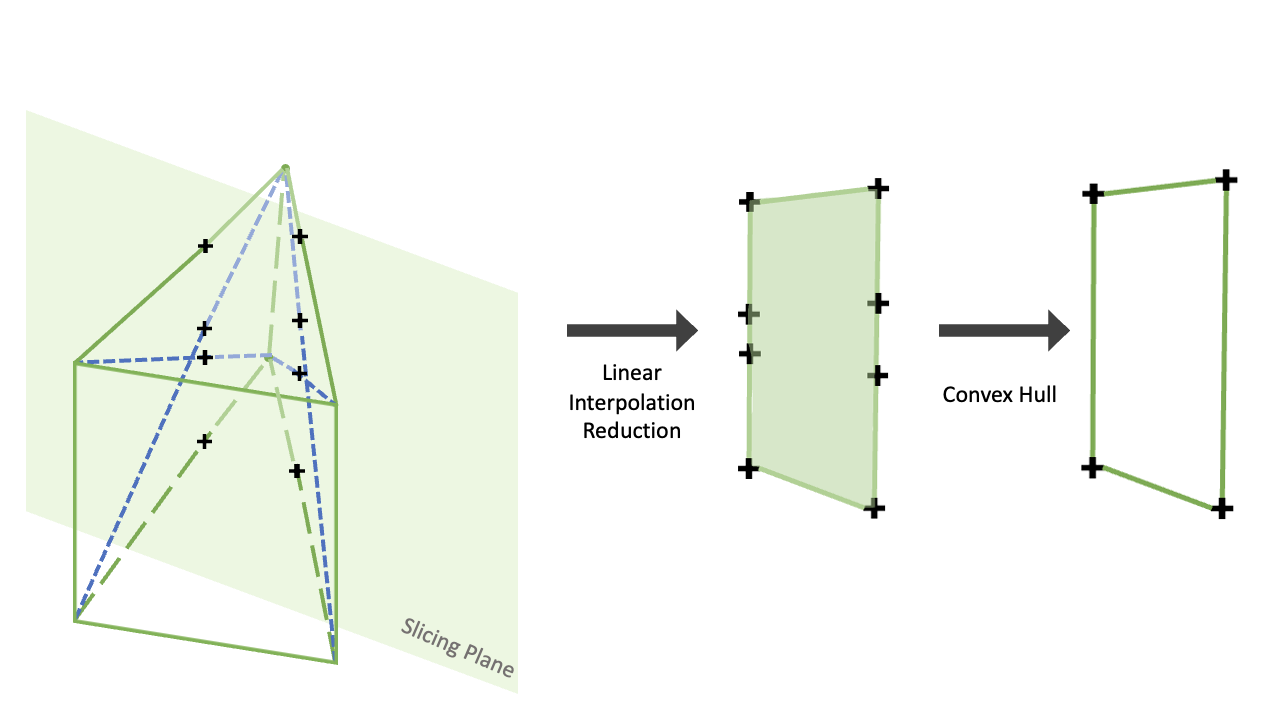}}
  \caption{2D and 3D examples of the Polytope slicing mechanism on ordered axes. }
  \label{slicing_step} 
\end{figure*}

 The actual slicing step is quite straightforward with the slicing mechanism merely consists of finding the intersection of a polytope with a hyperplane along a datacube axis. We first separate all vertices in the polytope into two separate groups, each group consisting of points on either side of the hyperplane. We then linearly interpolate between each pair of vertices where one vertex comes from one vertex group and the other from the other. We linearly interpolate these pairs to find the interpolated point which lies on the slice plane. Once we have done this for all pairs, we obtain a lower-dimensional polytope on the slice plane, which is in fact just the intersection of the original polytope with the slice plane, as wanted. This can be seen in Figure \ref{slicing_step} for some 2D and 3D examples.
\\
\noindent
As the original polytope is convex, this new intersection polytope is trivially also convex. As an optimisation step, we can thus take the convex hull of the intersection points at the end, using the QuickHull algorithm \cite{quickhull} for example. This does not change the lower-dimensional polytope because it is convex, but removes all interior vertices in its definition. As we slice high-dimensional polytopes, this can lead to major performance improvements. Indeed, without this last step, the number of vertex points in the polytope definition grows quadratically with each slice, which would considerably slow down the algorithm.
\subsubsection*{\textbf{Index Tree Construction}}
\begin{figure}[!ht]
         \centering
         \includegraphics[width=    0.8\linewidth]{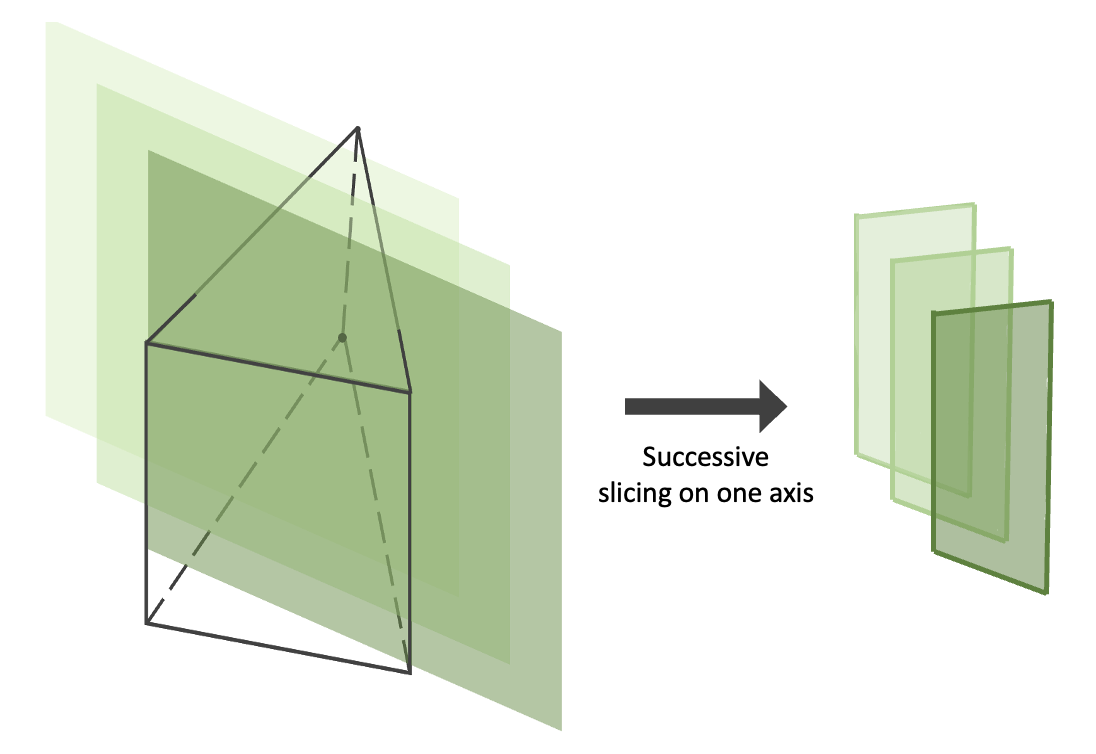}
         \caption{Successive slicing of a polytope along one axis at different indices, resulting in a list of lower-dimensional polytopes. }
         \label{sucessive_slicing}
\end{figure}

To ensure that we slice through all the requested polytopes defined on different axes of the datacube, we need to carefully keep track of which step in the extraction we are in. The way we achieve this in the Polytope extraction technique is to iteratively build an index tree.\\
We build the index tree by slicing along successive axes on the datacube one after another. For each axis of the datacube, we first find the polytopes defined on that axis. We then find the discrete indices on that axis contained within the extents of those polytopes and add them as children to the index tree. Next, we slice the necessary polytopes along each of the discrete datacube indices to obtain lower-dimensional polytopes. As shown in Figure \ref{sucessive_slicing}, these lower-dimensional polytopes are the intersection of the higher-dimensional polytopes with each of the axis indices slice hyperplanes. These new polytopes are the next polytopes we would like to now extract from the datacube. The algorithm therefore continues as before on these lower-dimensional polytopes if they exist. This process is re-iterated in Algorithm \ref{extraction_algo}.
\\
\noindent
Note that this works well on ordered axes. On categorical axes however, the slicing step is ill-defined as interpolation between indices is not possible. Nevertheless, recall that polytopes defined on categorical axes are in fact 1D points and thus instead of slicing, we only need to check whether those points exist in the datacube. Indeed, slicing does not matter in this case as the points are 1 dimensional and slicing, if it were well-defined, would therefore not produce any lower-dimensional polytopes anyway. We thus conclude that the process for constructing index trees presented in Algorithm \ref{extraction_algo} does in fact work well for categorical axes as well. \\
Algorithm \ref{extraction_algo} implies that we construct the index tree breath-first (layer by layer), instead of depth-first (constructing branches one after the other). This approach ensures that the algorithm does not loose track of what values inside the requested polytopes have already been found. It thus ensures that users get back all the points that are contained in the shape they requested.
\begin{algorithm*}[ht!]
\caption{Polytope Slicing Algorithm}
    \label{extraction_algo}
    \begin{algorithmic}[1]
    \State \textbf{Input:} list of polytopes $\mathcal{P}$
    \State Remove duplicate points in polytopes
    \For{axis in datacube axes}
        \State Find polytopes in $\mathcal{P}$ defined on axis
        \For{polytope in found polytopes}
            \State Find extents of these polytopes on axis
            \State Find discrete indices between extents from datacube 
            \State Add discrete indices as children to index tree
            \If{ axis is a categorical axis}
                \State Skip
            \ElsIf{axis is an ordered axis}
                \For{axis index in discrete indices}
                    \State Slice polytope along axis index to get lower-dimensional polytope
                \EndFor
            \EndIf                
        \EndFor
        \State{Update list of polytopes $\mathcal{P}$ by list of sliced lower-dimensional polytopes}
    \EndFor
    \end{algorithmic}
\end{algorithm*}

\section{Applications}
The Polytope data extraction algorithm has a wide range of interesting applications, from meteorology to healthcare. In this section, we first introduce the different Polytope interface levels before discussing some Polytope applications, describing specific examples and how Polytope has improved access to data in those cases. 
\subsection{Interface}
\begin{figure}[!ht]
         \centering
         \includegraphics[width=    1.0\linewidth]{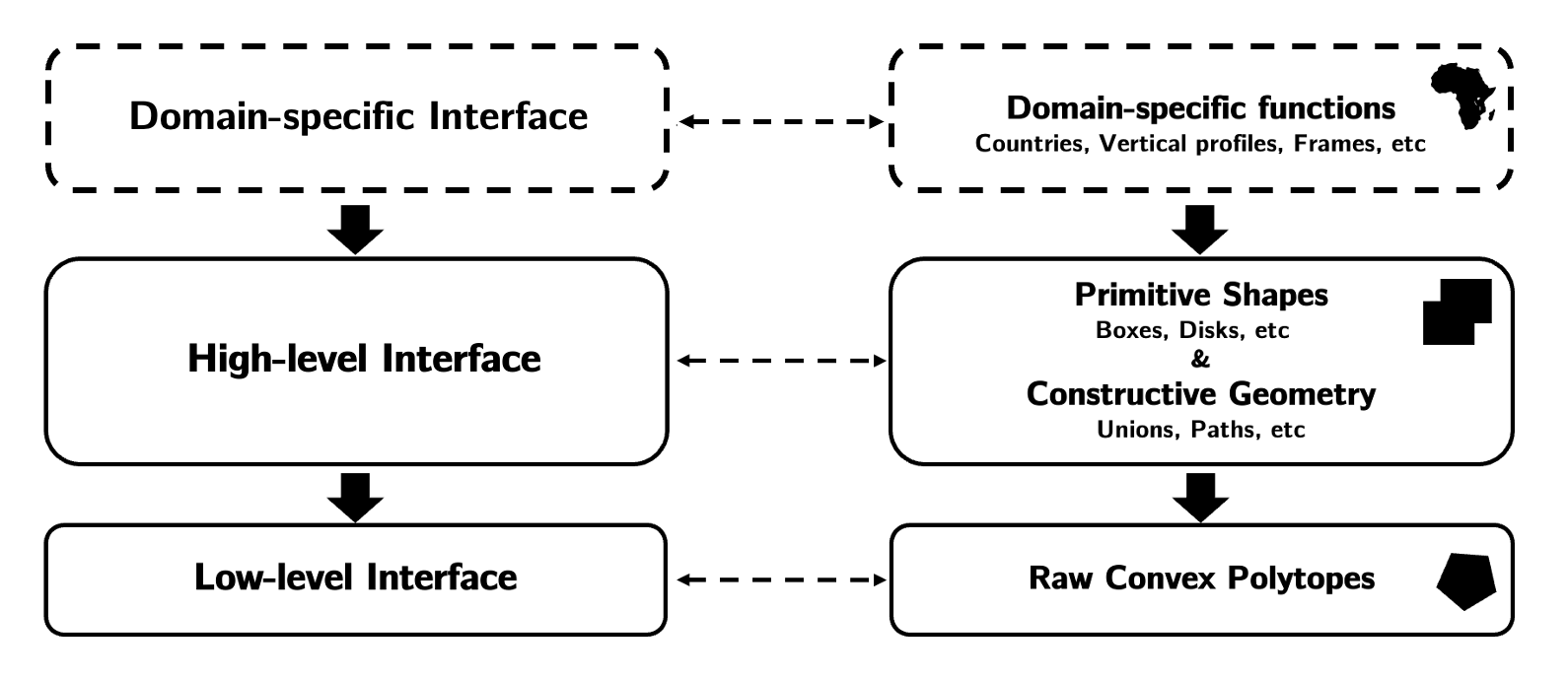}
         \caption{Polytope interface levels. }
         \label{apis}
\end{figure}

To facilitate interaction with the Polytope feature extraction algorithm, which only accepts polytopes as input, different interfaces can be implemented. 
The Polytope interfaces serve as platforms for the users to interact with the extraction algorithm. 
In particular, users will submit their request shapes and, after the algorithm has run, retrieve their desired data to and from these interfaces. 
To accommodate different types of users, several interface levels exist, which let users request a wide range of request shapes, from the low-level generic convex polytope to higher-level specialised requests. The two in-built low- and high-level Polytope interfaces are shown in Figure \ref{apis}. It is also possible to build domain-specific interfaces on top of these built-in interfaces, also shown in Figure \ref{apis}. Each level is built on top of another with the domain-specific interfaces using shapes from the high-level interface, which itself depends on the low-level interface.
\\
\noindent
These distinct interface levels are useful because depending on specific needs and familiarity with the Polytope extraction technique, users might want to request different types of shapes from our algorithm. \\
Through the domain-specific interfaces, users can request domain-specific functions. For example, a meteorological interface could be built to facilitate access to time-series, trajectories or country extraction, similar to the OGC EDR\cite{ogc-edr} standard.\\ 
Through the built-in high-level interface, users can request primitive shapes, such as disks or boxes, and then use constructive geometry operations, such as taking unions or sweeping along a path, to build more complicated shapes. Finally, through the low-level interface, users can directly provide a list of convex $n$-dimensional polytopes, specified by a list of their vertices. \\
Each level is built on top of its lower-level counterpart, so that shapes in a higher level are always defined by shapes in one of the lower levels. This implies that shapes in any of the interface levels are in fact always defined as a combination of convex low-level polytopes. These low-level polytopes are the building blocks of all possible Polytope requests. The interface is responsible for decomposing all user request shapes into these base convex polytopes. In the rest of the software, we can then work only on these convex polytopes and take a unified approach towards slicing any user request shape. 
\subsection{Polytope Use Cases}
We now describe some examples of how the Polytope feature extraction algorithm can be used in the fields of meteorology and healthcare.
\subsubsection*{\textbf{Meteorology}}
\begin{figure*} 
\captionsetup[subfigure]{font=footnotesize}
    \centering
  \subfloat[Timeseries\label{timeseries}]{
       \includegraphics[width=0.4\linewidth]{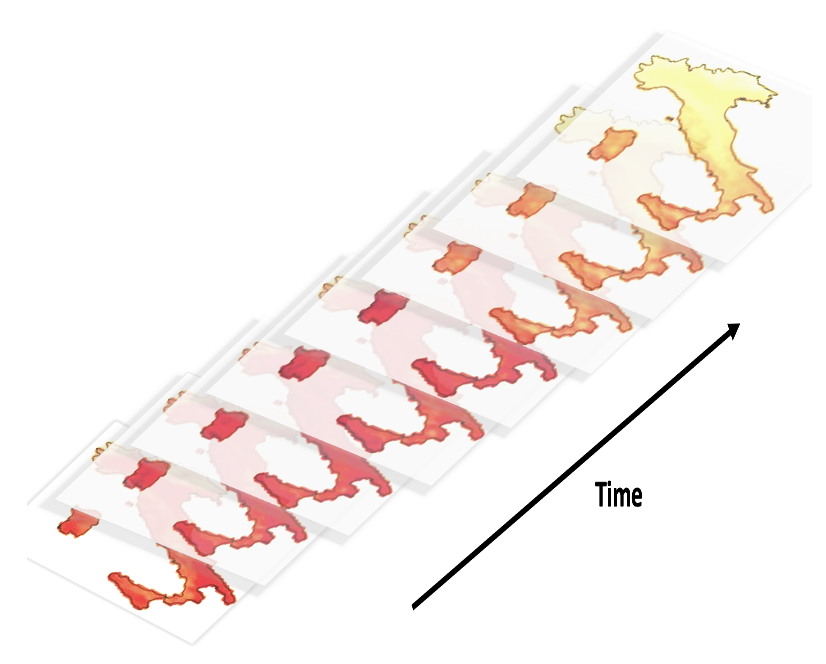}}
    \hfill
  \subfloat[Flight Path\label{3d}]{
        \includegraphics[width=0.55\linewidth]{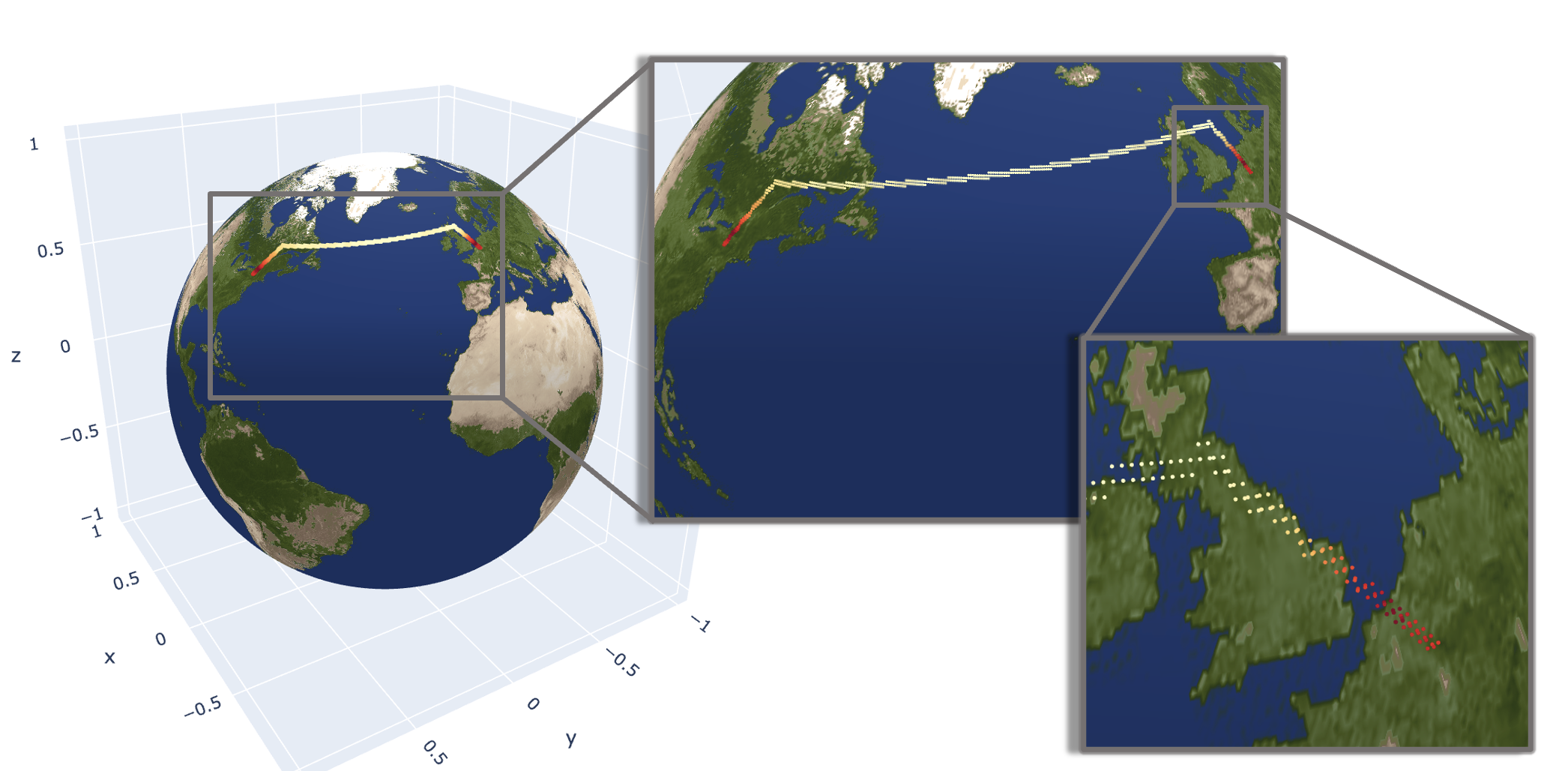}}
  \caption{Meteorology examples, where the coloured points are all the points that were extracted using the Polytope algorithm.}
  \label{polytope_examples} 
\end{figure*}

At the European Center for Medium-Range Weather Forecasts (ECMWF), about 300 TiB of numerical weather prediction (NWP) data are produced daily. This data is very high-dimensional and is usually represented as a datacube of 7 or 8 dimensions depending on the forecast type. 
Over the next few years, following the pioneering work of the Destination Earth initiative \cite{destine} with planned resolution increases in the weather model, data production will grow to about a petabyte of data a day. \\
The current data extraction mechanism implemented at ECMWF is one of the traditional bounding box approaches. When a user wants to extract data on a country for example, they would need to send a request for a bounding box around that country. Moreover, the current extraction technique requires either full data fields or at very best bounding boxes of data fields to be read from the system even when users only request a smaller portion of data. With future petabyte-scale datacubes, this approach will become impractical, especially when trying to accommodate for thousands of users. 
The Polytope extraction technique helps alleviate many of the challenges faced by the system in this case. 
It makes returning data to users much more efficient because only the required bytes are read from the I/O system. \\
Below, we provide a few practical examples and use cases where Polytope might help meteorological data users extract data more efficiently.
\paragraph*{Timeseries}
Imagine a user interested in extracting the temperature over Italy for the next two weeks. 
She would currently have to transpose the temperature fields along the time axis to be able to then individually extract each temperature field at a given timestep. For each timestep, she would have to cut the shape of Italy from the bounding box she retrieved before finally getting the exact data she wanted. With the Polytope extraction technique, she can instead directly request the timeseries over Italy and get back only the precise bytes she is interested in, as shown in Figure \ref{timeseries}. Note that compared to the 3D bounding box the user would currently retrieve, we see a data reduction of more than {73\%} when using Polytope. Furthermore, note that  meteorological  data  users  are  usually  more  interested in extracting data over particular cities or specific points in space rather than  whole  regions.  However,  since  users  currently first have to transpose their data and then retrieve bounding  boxes  around  their  locations  of  interest,  in  most  cases they  directly  extract  data  over  broader  regions  than  just the  specific  locations  they  would  like  to  access.  With the Polytope algorithm, complicated pre-processing manipulations before extraction are not needed anymore and users only retrieve the relevant timeseries data from the datacube.
\paragraph*{Flight Path}
Now, imagine a user interested in the flight conditions over his plane journey from Paris to New York. Using the current extraction technique, he would get back a 4 dimensional box over 3D space and time, containing much more data than what he is interested in. With the Polytope extraction technique, he will instead only get back the specific points he is interested in in the datacube without any need for post-processing, as shown in Figure \ref{polytope_examples}b. Note that compared to the 4D box the user would currently get back, with Polytope, we experience a data reduction of more than {99.99\%}.
\\[10pt]
In both cases, the requested shapes are not axis-aligned and are therefore also not well approximated by bounding boxes. We thus see a significant data reduction when using the Polytope extraction technique compared to the traditional bounding box approach. Importantly, we observe that I/O is reduced when using the Polytope extraction algorithm. Moreover, using the Polytope algorithm is particularly useful for the users, who do not need to do any post-processing to their data in order to get their requested shape. 
\subsubsection*{\textbf{Healthcare}}
\begin{figure}[!ht]
         \centering         \includegraphics[width=0.8\linewidth]{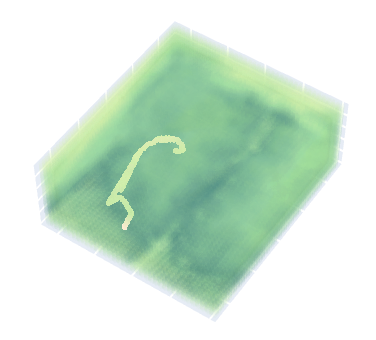}
         \caption{MRI scan example, where the white line is an extracted blood vessel, reaching from the cavernous segment of the internal carotid artery to the end of the middle cerebral artery.}
         \label{mri}
        \label{egs}
\end{figure}

Similarly to the weather forecasting industry, the healthcare industry faces complex data handling challenges. Already in 2019, \cite{wef_healthcare} estimated hospitals to generate tens of petabytes of data a year. As discussed in the previous example, working on this amount of data is extremely difficult and a tool like the Polytope algorithm could significantly help alleviate much of the difficulty involved. 
\\
\noindent
A particular example of how Polytope can be used in the healthcare field is provided below.
\paragraph*{MRI Blood Vessel Detection}
A clinically relevant application of MRI is the detection and characterization of plaque formation in (potential) stroke patients. This requires high-resolution scans using multiple MRI contrast weighting to comprehensively characterize the size and composition of plaque components. Using current extraction techniques, a clinician would have to download multiple entire MRI scan and then manually extract and compare the relevant data of interest from each of those scans. With the Polytope extraction technique however, it is possible to directly extract the required multi-contrast blood vessel data without further delay or expensive post-processing work, as is shown in Figure \ref{mri} for a single high resolution black-blood vessel wall MRI dataset \cite{mri2, mri1}.
\section{Performance and Scalability}
Polytope is predicted to considerably decrease the computational cost of extracting non-orthogonal data from hypercubes. In this section, we justify this claim by first analysing the performance of the Polytope algorithm and then investigating the data reductions achieved on practical use cases when using the Polytope algorithm instead of traditional extraction methods. We conclude this section by discussing these results and their significance.
\subsection{Performance}
\begin{figure*} 
\captionsetup[subfigure]{font=footnotesize}
    \centering
  \subfloat[ \label{fig1}]{%
       \includegraphics[width=0.5\linewidth]{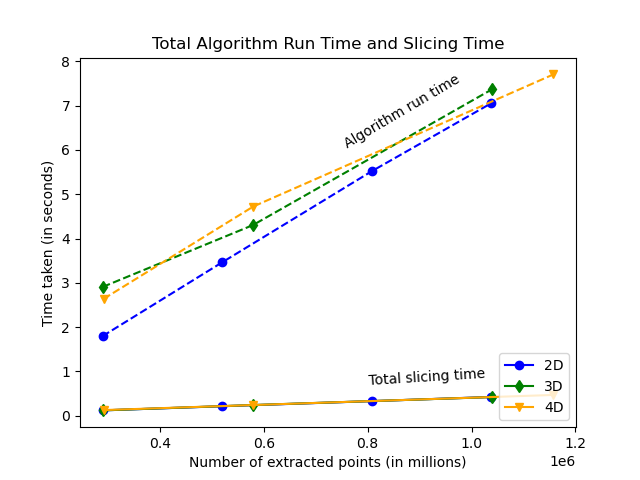}}
    \hfill
  \subfloat[\label{fig2}]{%
        \includegraphics[width=0.5\linewidth]{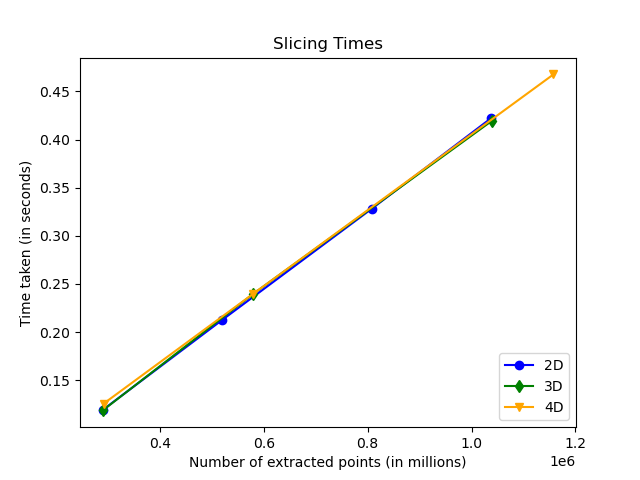}}
    \\
      \subfloat[\label{fig3}]{%
       \includegraphics[width=0.5\linewidth]{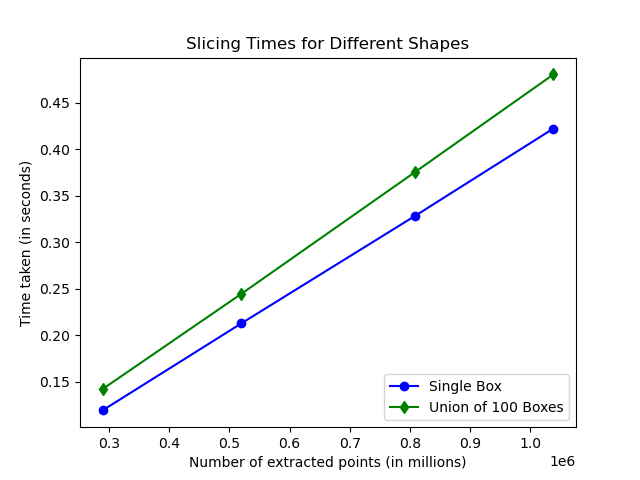}}
    \hfill
  \subfloat[\label{fig4}]{%
        \includegraphics[width=0.5\linewidth]{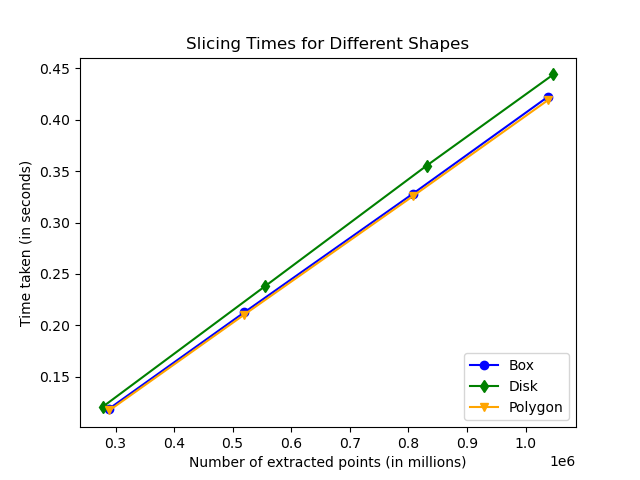}}
  \caption{Performance plots of the time taken to run the Polytope algorithm in function of the number of points extracted by the algorithm. In (a), we plot both the total algorithm run time (dashed lines) and the slicing time (solid lines) for different dimensions. In (b), we plot the slicing times for different dimensions. In (c) and (d), we plot the slicing times for different shape types. Algorithm timings on an Apple M1 Pro chip (3.2 GHz, 8 cores) and 16GB DDR5.}
  \label{performance_fig} 
\end{figure*}

There are many factors impacting the performance of the Polytope algorithm. To characterise it better, we identified some of the key features affecting how long the Polytope algorithm takes to extract points from datacubes: the number of extracted points, the dimension of the input shape and its geometry or how the input shape was constructed by the user. 
\\
\noindent
Consider two new time quantities, the total slicing time and the algorithm run time. The slicing time is the total accumulated time spent just slicing, without constructing the index tree. The total algorithm run time however is the time it takes to perform all of Algorithm \ref{extraction_algo}, including both the slicing time as well the time spent constructing the whole index tree.
In Figure \ref{performance_fig}, we have plotted both of these time quantities in different settings. In each subplot, we have varied one of the previously identified feature and kept all others constant. This lets us gain an understanding of how each individual feature influences the performance of the Polytope algorithm. Note also that we have not included the time taken in I/O to fetch the data from storage as this depends on the storage medium we use and is not strictly part of the Polytope algorithm.
\\
\noindent 
In Figure \ref{fig1}, we first plot both the slicing time as well as the total algorithm run time for request shapes of different dimensions. We observe that the dimension of the shape does not significantly impact the algorithm's performance. This is due to the fact that, even when slicing higher-dimensional shapes, the Polytope algorithm spends most of its time slicing lower-dimensional polytopes. In fact, note that the number of polytopes to process at each step in the algorithm quickly grows every time we slice to a lower dimension. 
For example, imagine we slice a 4D box which contains two indices on each dimension.  We first have to perform 2 4D slices. This then gives us 2 3D boxes, which we now have to slice. Each of these 3D boxes still has two indices on each dimension along which we need to slice. For each 3D box, we thus have to perform 2 3D slices. Considering there are 2 3D boxes, this implies we have to perform 4 3D slices in total. If we continue this logic, at the end of the algorithm, we will have performed 2 4D slices, 4 3D slices, 8 2D slices and 16 1D slices. This illustrates why the lower-dimensional slices do in fact take up most of the algorithm's slicing time. \\
\noindent
Furthermore, we note in Figure \ref{fig1} that the slicing time is much lower than the total algorithm run time.
This is because Polytope is currently using the XArray \cite{xarray} library for datacube implementation, and relies on XArray to look up discrete axis indices on the datacube. This is a step we believe still needs to be optimised by developing more efficient datacube look-up mechanisms and alternative datacube implementations. Meanwhile, in Figures \ref{fig2}-\ref{fig4}, we use the slicing time rather than the total algorithm run time to estimate Polytope's performance, thus excluding this dependency.\\
\noindent 
Figure \ref{fig2} shows the behaviour of the slicing time in more detail. In particular, we notice that like the total algorithm run time in Figure \ref{fig1}, the slicing time does not seem to depend on the dimension of the input shape. We also observe that the slicing time grows linearly with the number of datacube points the algorithm finds in the input shape. As discussed before, this is due to the fact that most of the slicing time is spent performing 1D slices. Indeed, increasing the number of points contained in the shape is effectively equivalent to increasing the number of 1D slices to perform to find those points. As it is those slices that make up most of the slicing time anyway, it is thus natural that the performance of the algorithm grows linearly with the number of points contained in the shape.\\
\noindent
In Figures \ref{fig3} and \ref{fig4}, we now investigate the impact of the input shape's geometry and how it was constructed by the user.\\\noindent 
In Figure \ref{fig3}, we first study how constructing a shape by taking a union of smaller sub-shapes affects the performance of the algorithm compared to directly specifying the input shape as one single object in 2 dimensions. We see that the performance when the shape is constructed using unions is worse than when the shape is specified as one single object. This is due to the fact that when we request shapes as unions of sub-shapes, we first slice each sub-shape individually in the algorithm before combining the results of these steps into one single output. Because we first slice each sub-shape individually, we in fact slice along all the sub-shapes edges. As the sub-shapes touch along their edges, we thus end up slicing along the edges several times, which increases the slicing time compared to when this does not happen in the non-union shape case. This is relevant where the input geometry has been produced via triangulation or mesh generation.\\\noindent
In Figure \ref{fig4}, we finally analyse how Polytope's different 2 dimensional high-level API primitive shapes: the box, disk and polygon shapes, influence the algorithm's performance. Here, we observe that the box and polygon shapes perform similarly while the disk shape has a slightly worse performance than the other two shapes. Because the polygon shape we inputted is actually a square, we can conclude from this observation that the algorithm's performance is mostly impacted by the number of vertices of the shape, as well as how ``non-orthogonal" it is. In particular, if the shape has many edges cutting across some axes, then it is less likely that there will be duplicates when we compute the intersection points in the slicing step. We will thus then have to perform more slicing later on in the algorithm. The same logic holds if there are more vertices in the shape, which also increases the number of intersection points computed in the slicing step.  
\subsection{Bound on Number of Slices}
As we just discussed, the time quantity of interest to us in evaluating the performance of the Polytope algorithm is the slicing time. The slicing time largely depends on a related quantity which is the number of slices performed during the algorithm. In this subsection, we quickly determine a theoretical upper bound to this related quantity.\\
Suppose we query an $m$ dimensional shape using Polytope and let $n_i$ be the maximal number of discrete indices stored along each of the $i=1, \dots, m$ axes of the datacube which are contained within the requested shape. Since we do not know a priori exactly how convex or non-box-like the requested shape is, we have to assume the worst-case scenario that the shape is in fact a box. \\
To find the datacube points within that worst-case box shape, the Polytope algorithm now first has to slice $n_1$ times along the first axis dimension. This produces $n_1$ $(m-1)$-dimensional box shapes. We then have to slice each of these lower-dimensional box shapes $n_2$ times along the second dimension. This creates an additional $n_1 \times n_2$ slices.\\
If we continue this process up to the last 1D slices, we finally see that the number of slices performed during the Polytope algorithm, $N_{slices}$ is bounded by 
\begin{align*}
    N_{slices} \leq n_1 + &n_1 \times n_2 + \dots + n_1\times \dots\times n_m\\
    &= \sum_{i=1}^m \,\prod_{j=1}^i \, n_j\,.\end{align*}
We see that, as expected, this upper bound is dominated by the number $ \prod_{i=1}^m n_i$ of 1D slices, which will take up most of the slicing time. 
As we saw in the previous subsection however, 1D slices are relatively inexpensive to perform and in all of the examples in Figure \ref{performance_fig}, the slicing time remains under a second. 
\subsection{Data Reductions}
\begin{table*}
\renewcommand{\arraystretch}{1.3}
\caption{Polytope Data Reduction and Performance. Note that for completeness, both slicing and total algorithm run time are included, although the slicing time is more indicative of the Polytope algorithm performance, as discussed in subsection 5.1. \\Algorithm timings on an Apple M1 Pro chip (3.2 GHz, 8 cores) and 16GB DDR5.}
\label{table}
\centering
\begin{tabular}{ |C{1.8cm}||C{1.8cm}|C{1.8cm}|C{1.8cm}||C{1.8cm}|C{1.8cm}||C{1.8cm}|C{1.8cm}| }
 \hline
 \textbf{Example Shape}& \textbf{Data retrieved with traditional approach}& \textbf{Data retrieved with bounding box approach} & \textbf{Data retrieved with Polytope algorithm}&\textbf{Reduction factor compared to traditional approach} &\textbf{Reduction factor compared to bounding box approach}& \textbf{Slicing time}&\textbf{Total algorithm run time}\\
 \hline
  \hline
  Box around Germany & 50.4 MB & 44 KB& 44 KB&1173 $\times$&1 $\times$&2.3e-3 s&0.03 s\\
  \hline
  Timeseries of London over 14 days  &5.5 GB& 896 B & 896 B& 6591049 $\times$&1 $\times$& 1.4e-4 s&0.13 s\\
  \hline
     Vertical Profile over 20 layers - Rome &1 GB&  800 B & 800 B & 1342177 $\times$ & 1 $\times$& 4.6e-5 s&0.02 s\\
  \hline
  \hline
 Country shape of France   &50.4 MB& 67.7 KB   & 32.3 KB &  1598 $\times$&2 $\times$&0.03 s&0.94 s\\
  \hline
 Country shape of Norway &50.4 MB&  171.4 KB & 29.9 KB   & 1726 $\times$& 6 $\times$ &0.06 s&1.97 s\\
  \hline
 Flight Path from Paris to New York &7.9 GB& 247.3 MB & 4.9 KB & 1690561 $\times$ &51681 $\times$&0.07 s&0.18 s\\
 \hline
 MRI Blood Vessel &1 GB&   1.5 MB  & 4.5 KB &233017 $\times$&341 $\times$&0.10 s&0.35 s\\
 \hline
\end{tabular}
\end{table*}
\noindent
Although the Polytope algorithm represents an additional step to perform before extracting the data and it might thus at first glance seem like it has a much higher time complexity than traditional extraction approaches, it is important to remember the true purpose of the Polytope algorithm.
Polytope is a tool which computes the precise bytes of data a user wants to access. Using this tool therefore implies that users only extract exactly the data points they need, which significantly reduces the number of points to be read from the I/O system compared to the alternative "bounding box" extraction techniques. The exact data reduction statistics for the examples mentioned in the previous section are shown in Table \ref{table}. \\
In Table \ref{table}, the first 3 columns show the number of bytes retrieved when using different extraction techniques. In particular, note the clear distinction between the first two columns which differentiate the bounding box approach described earlier from the state of practice extraction methods taken in the fields of meteorology and healthcare respectively, which are even less optimal than the bounding box approach. Indeed, it is important to note here that one of the widely used extraction approaches in the field of meteorology for example is to extract whole fields, which are 2D arrays of latitude and longitude around the whole globe, from datacubes. Similarly, in the field of healthcare, MRI scans are currently stored as 3D images. The bounding box approach is thus already a clear improvement compared to these approaches. As we see in Table \ref{table} however, Polytope performs even better than the bounding box approach. This can be clearly observed in the fifth column, where we provide the reduction factor of the data retrieved when using the Polytope algorithm compared to the bounding box approach. The sixth column shows the total reduction of the data retrieved when using the Polytope algorithm compared to the state of practice extraction methods taken in the meteorology and healthcare fields. The final two columns then show the two slicing and total algorithm run times discussed above for each of our example shapes.  \\
Along the rows, we differentiate between different types of shapes. On the first 3 rows, we first test the Polytope algorithm on shapes that are defined orthogonally along their axis and which could be directly extracted using the bounding box approach. For these 3 rows, as we see in the fifth column, using the Polytope algorithm instead of the bounding box approach does not reduce the size of the retrieved data further. Note however that running the Polytope algorithm in these three examples does not take significant time. In the latter 4 rows, we then experiment using the Polytope algorithm to retrieve more complicated non-orthogonal or axis-aligned shapes. Already for country shapes in 2D, we see that there is a significant data reduction when using the Polytope algorithm compared to the bounding box approach, with a reduction factor of up to 6 times in some cases. When considering higher dimensional shapes, and especially ``path"-like shapes such as flight paths, we experience an even higher reduction factor. Indeed, in the 4D case of the flight path from Paris to New York mentioned above, about 350 times less data is returned to the users when using the Polytope algorithm instead of the bounding box approach. Again, note here that in most examples, the Polytope algorithm takes below a second to run whilst reducing the retrieved data size by a factor of at least 1000 compared to the traditional approaches.\\
Importantly here, note that Polytope is able to perform the exact same orthogonal extractions as the bounding box approach in minimal time, whilst significantly outperforming the bounding box approach when extracting more complicated shapes. This suggests that the Polytope algorithm performs at least as well as the bounding box approach and thus makes it a strong competitor to this approach.
\noindent  
\subsection{Discussion}
Note that in subsection 5.1 above, in Figure \ref{fig1}, we did not include the time spent extracting data from the datacube. This is because this data extraction time is very dependent on the storage medium on which the algorithm is run. We expect that the cost of performing the Polytope algorithm will be significantly less than the savings made by retrieving less data. In particular, we suspect that hardware that supports high performance random-read, such as flash based devices for example, will benefit massively from the Polytope algorithm. \\
Compared to traditional methods, the Polytope algorithm is an additional step in the extraction process which takes time to run. However, as we saw in this section, and in Figure \ref{performance_fig} especially, the Polytope algorithm is efficient and scalable, being able to locate more than a million points in less than half a second. Moreover, as already mentioned, when discussing performance of the algorithm, it is especially important to also consider its wider role in the total extraction process. As we saw in Table \ref{table}, the Polytope algorithm allows users to extract much less data than they would have done using a more traditional approach. As reading and returning data is usually a costly operation, it implies that, when incorporated in a complete extraction pipeline, the Polytope algorithm will make data extraction more efficient than the current state of practice. The slightly more expensive slicing mechanism inside the Polytope algorithm will thus be outweighted by the actual performance improvement of the whole data extraction pipeline. \\
Before being able to  quantify the true performance and benefits of the Polytope algorithm, we will therefore need to perform a more in-depth analysis of its behaviour within a complete data extraction framework. This more detailed analysis is the subject of ongoing work and material for a future communication.
\section{Conclusion}
In this paper, we introduced a new data extraction algorithm called Polytope, which has the capability of extracting arbitrary geometrical shapes from a datacube. This new technique allows users to directly compute the precise bytes of interest to them before requesting these bytes from a datacube. This approach leads to many benefits, both for the users and the data providers. For data providers, much less I/O is needed whereas for users, the need for further post-processing after extraction is alleviated. We described the structure of this novel extraction algorithm and explained in more detail some of its key features. We then showed a few use cases of the Polytope extraction technique before finally analysing the performance of this method. 
Future steps include performing a more in-depth analysis of the algorithm performance, as well as a rigorous discussion of Polytope's use cases in different scientific fields. 

\section*{Acknowledgments}
\addcontentsline{toc}{section}{Acknowledgment}
This work is an important contribution to, and is funded by, the EU’s Destination Earth initiative. The authors would also like to thank Matthijs de Buck and the Nuffield Department of Clinical Neurosciences at the University of Oxford for providing data for the MRI Blood Vessel Detection use case.

\bibliographystyle{IEEEtran}

\bibliography{IEEEabrv, bibfile} 

\begin{thebibliography}{10}
\providecommand{\url}[1]{#1}
\csname url@samestyle\endcsname
\providecommand{\newblock}{\relax}
\providecommand{\bibinfo}[2]{#2}
\providecommand{\BIBentrySTDinterwordspacing}{\spaceskip=0pt\relax}
\providecommand{\BIBentryALTinterwordstretchfactor}{4}
\providecommand{\BIBentryALTinterwordspacing}{\spaceskip=\fontdimen2\font plus
\BIBentryALTinterwordstretchfactor\fontdimen3\font minus
  \fontdimen4\font\relax}
\providecommand{\BIBforeignlanguage}[2]{{%
\expandafter\ifx\csname l@#1\endcsname\relax
\typeout{** WARNING: IEEEtran.bst: No hyphenation pattern has been}%
\typeout{** loaded for the language `#1'. Using the pattern for}%
\typeout{** the default language instead.}%
\else
\language=\csname l@#1\endcsname
\fi
#2}}
\providecommand{\BIBdecl}{\relax}
\BIBdecl

\bibitem{datacube}
J.~Gray, S.~Chaudhuri, A.~Bosworth, A.~Layman, D.~Reichart, M.~Venkatrao,
  F.~Pellow, and H.~Pirahesh, ``{Data cube: A relational aggregation operator
  generalizing group-by, cross-tab, and sub-totals},'' \emph{Data mining and
  knowledge discovery}, vol.~1, pp. 29--53, 1997.

\bibitem{matlab}
D.~J. Higham and N.~J. Higham, \emph{{MATLAB guide}}.\hskip 1em plus 0.5em
  minus 0.4em\relax SIAM, 2016.

\bibitem{xarray}
S.~Hoyer and J.~Hamman, ``{xarray: ND labeled arrays and datasets in Python},''
  \emph{Journal of Open Research Software}, vol.~5, no.~1, 2017.

\bibitem{xtensor}
{Xtensor Stack}, ``{Xtensor Documentation},'' In \emph{Xtensor} (Version
  0.24.6). Retrieved from Read the Docs:
  \url{[https://xtensor.readthedocs.io/en/latest/]}, 2023.

\bibitem{sql}
J.~Melton and A.~R. Simon, \emph{{SQL: 1999: understanding relational language
  components}}.\hskip 1em plus 0.5em minus 0.4em\relax Elsevier, 2001.

\bibitem{tquel}
R.~Snodgrass, ``{The temporal query language TQuel},'' \emph{ACM Transactions
  on Database Systems (TODS)}, vol.~12, no.~2, pp. 247--298, 1987.

\bibitem{olap}
K.~Morfonios, S.~Konakas, Y.~Ioannidis, and N.~Kotsis, ``{ROLAP implementations
  of the data cube},'' \emph{ACM Computing Surveys (CSUR)}, vol.~39, no.~4, pp.
  12--es, 2007.

\bibitem{polytopedef}
P.~Wolfe, ``Finding the nearest point in a polytope,'' \emph{Mathematical
  Programming}, vol.~11, pp. 128--149, 1976.

\bibitem{mesh}
S.~J. Owen, ``A survey of unstructured mesh generation technology,''
  \emph{IMR}, vol. 239, p. 267, 1998.

\bibitem{quickhull}
C.~B. Barber, D.~P. Dobkin, and H.~Huhdanpaa, ``{The QuickHull algorithm for
  convex hulls},'' \emph{ACM Transactions on Mathematical Software (TOMS)},
  vol.~22, no.~4, pp. 469--483, 1996.

\bibitem{ogc-edr}
M.~Burgoyne, D.~Blodgett, C.~Heazel, and C.~Little, ``{OGC API-Environmental
  Data Retrieval Standard},'' \emph{Open Geospatial Consortium Inc., Wayland,
  MA, USA, OpenGIS{\textregistered} Implementation Specification OGC}.

\bibitem{destine}
N.~Wedi, T.~Quintino, U.~Modigliani, V.~Baousis, T.~Geenen, I.~Sandu, P.~Bauer,
  J.~Hoffmann, and D.~Thiemert, ``{Destination Earth: Digital Twins of the
  Earth System},'' Copernicus Meetings, Tech. Rep., 2022.

\bibitem{wef_healthcare}
\BIBentryALTinterwordspacing
{World Economic Forum}, ``{4 ways data is improving healthcare},'' 2019,
  accessed on April 24 2023. [Online]. Available:
  \url{https://www.weforum.org/agenda/2019/12/four-ways-data-is-improving-healthcare}
\BIBentrySTDinterwordspacing

\bibitem{mri2}
O.~Viessmann, L.~Li, P.~Benjamin, and P.~Jezzard, ``T2-weighted intracranial
  vessel wall imaging at 7 tesla using a dante-prepared variable flip angle
  turbo spin echo readout (dante-space),'' \emph{Magnetic resonance in
  medicine}, vol.~77, no.~2, pp. 655--663, 2017.

\bibitem{mri1}
M.~H. de~Buck, J.~L. Kent, A.~T. Hess, and P.~Jezzard, ``Parallel transmit
  dante-space for improved black-blood signal suppression at 7 tesla,'' in
  \emph{Proceedings of the 31st Annual Meeting of ISMRM}, vol. 2092, 2022.

\end{thebibliography}

\end{document}